# Tunable Magnets: modeling and validation for dynamic and precision applications

Silvan G. Viëtor, Jo W. Spronck, S. Hassan HosseinNia

*Department of Precision and Microsystems Engineering (PME)*
*Faculty of Mechanical, Maritime and Materials Engineering (3mE)*
*Delft University of Technology*

s.g.vietor@student.tudelft.nl

**Abstract**

Actuator self-heating limits the achievable force and can cause unwanted structural deformations. This is especially apparent in quasi-static actuation systems that require the actuator to maintain a stable position over an extended period. As a solution, we use the concept of a Tunable Magnet. Tunable magnets rely on in-situ magnetization state tuning of AlNico to create an infinitely adjustable magnetic flux. They consist of an AlNiCo low coercivity permanent magnet together with a magnetizing coil. After tuning, the AlNiCo retains its magnetic field without further energy input, which eliminates the static heat dissipation. To enable implementation in actuation systems, the AlNiCo needs to be robustly tunable in the presence of a varying system air-gap. We achieve this by implementing a magnetization state tuning method, based on a magnetic circuit model of the actuator, measured AlNiCo *BH* data and air-gap flux feedback control. The proposed tuning method consists of 2 main steps. The prediction step, during which the required magnet operating point is determined, and the demagnetization step, where a feedback controller drives a demagnetization current to approach this operating point. With this method implemented for an AlNiCo 5 tunable magnet in a reluctance actuator configuration, we achieve tuning with a maximum error of 15.86 mT and a minimum precision of 0.67 mT over an air-gap range of 200 μm. With this tuning accuracy, actuator heating during static periods is almost eliminated. Only a small bias current is needed to compensate for the tuning error.

## 1. Introduction

One of the challenges in modern precision actuation systems is thermal stability. Heat dissipated in the actuator coils can cause unwanted structural deformations and therefore limits the achievable force and accuracy of the system [1][2]. This is especially a concern in quasi-static actuation systems, where the actuator needs to maintain a stable stationary position over an extended period. During this time the actuator is still dissipating heat to produce the required static force. In this paper, we use a concept called a *Tunable Magnet* (TM), which has the potential to increase energy efficiency and therefore thermal stability in precision actuation systems. A TM consists of an AlNiCo low coercivity permanent magnet (PM) combined with a coil to in-situ adjust its magnetization. After magnetizing, the AlNiCo retains its magnetic field without further energy input. TMs can therefore significantly reduce the constant heat dissipation during stationary periods when used in actuators. Promising applications include set-and-forget alignment systems, adjustable magnetic gravity compensators [2][3] and highly stable microscopy stages.

The idea of using in-situ magnetization adjustment of AlNiCo has been the subject of prior research. In 2010, A.N. Knaian [4] introduced the Electropermanent Magnet. It consists of an AlNiCo 5 magnet in parallel with a NdFeB magnet, together with a magnetizing coil. Changing the polarization of the low coercivity AlNiCo magnet with a pulsed external field effectively switches the magnet assembly 'On' and 'Off'. Devices that rely on in-situ tuning, as opposed to switching, of magnetization of AlNiCo have been realized for magnetic clamping applications [5-7], flux weakening/boosting in PM motors [8-10] and force tuning in a magnetic gravity compensator [2][3].

The main challenge in implementing the TM concept is how to determine the correct coil current to tune the magnet to the desired strength. This magnetizing current depends on both the present magnetization state and the reluctance of the magnetic circuit in which the TM is placed. In actuation applications, this reluctance varies, usually due to a changing air-gap. Some research has been done on trying to predict the magnetizing current using models or measurements. For example, [9] and [10] base their estimate on a parallelogram-shaped hysteresis curve approximation, combined with a simple magnetic circuit model. However, small deviations due to model inaccuracy and geometry tolerances can lead to substantial differences in resulting magnet strength, because the magnetization of the AlNiCo 5 magnet being used is very sensitive to variations in the applied field. To overcome this, [2] uses measurements of the AlNiCo $BH$ characteristic to determine the magnetizing current. By entirely demagnetizing before each tuning step, only a single measured curve is needed, because the starting magnetization condition is always identical. This leads to good repeatability at a fixed actuator position but attained magnetization values will still change with changing system air-gap.

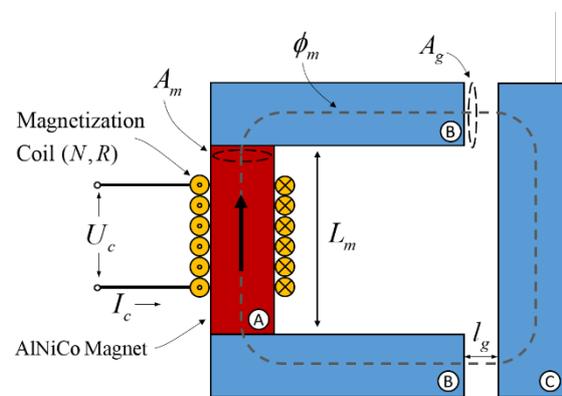

**Figure 1:** Schematic of Tunable Magnet actuator with AlNiCo magnet [A], soft steel pole pieces [B] and keeper bar [C].

This work aims to develop a TM that can be robustly tuned in the presence of a varying system reluctance, in order to enable its implementation in actuation applications. We achieve this by implementing a magnetization state tuning method based on measurements of AlNico $BH$ data, a magnetic circuit model, and air-gap flux feedback control. In the current implementation, we use AlNiCo 5, but the method presented can be used for other grades of AlNiCo and potentially other low coercivity PMs as well.

The rest of this paper is organized as follows: Section 2 describes the operation of PMs and derives the magnetic circuit model. In section 3 this model is used to develop a magnetization state tuning method. Part of this method is a demagnetization controller which is elaborated in section 4. The experimental setup and measurement results are reported in section 5. Section 6 concludes on these results and gives an outlook on the use of TMs in precision actuation systems.

## 2. Modeling and Permanent Magnet Operation

In this research, we assume that the *Tunable Magnet* is applied in a standard gap closing reluctance actuator topology, as shown in Fig 1. The system consists of the AlNiCo PM [A] with length $L_m$ and cross-section $A_m$ together with a magnetizing coil. The coil has $N$ number of turns which carry a current $I_c$ supplied by a voltage source $U_c$. The resulting coil resistance is denoted by $R$. Soft steel pole pieces [B] with cross section $A_g$ are on both sides of the magnet, and a soft steel keeper bar [C] completes the magnetic circuit. Applying conservation of flux and evaluating Ampere's Law over the contour $\phi_m$, gives the following lumped parameter model of the TM actuator [11]:

$$B_m A_m = k_1 B_g A_g \quad (1)$$
$$H_m L_m + 2 k_2 H_g l_g = N I_c \quad (2)$$
$$B_g = \mu_0 H_g \quad (3)$$

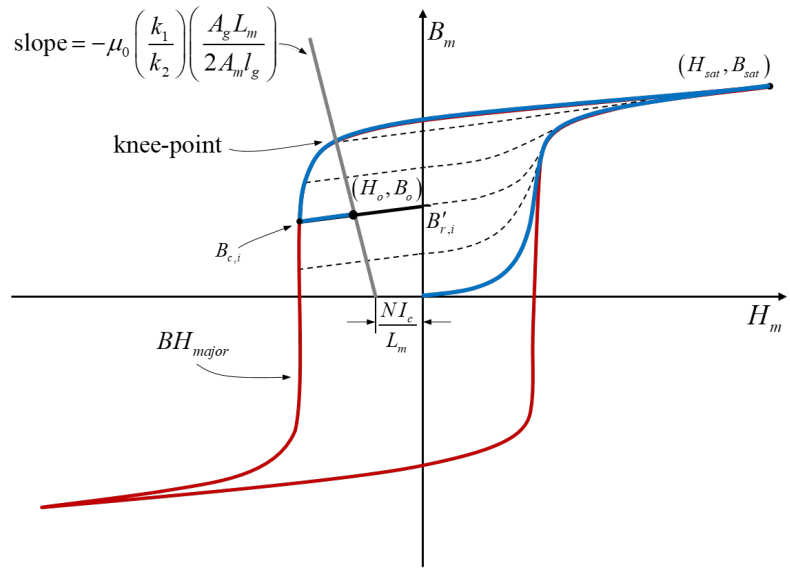

**Figure 2:** Major *BH* curve (red) with a load-line (grey) and operating point $(H_o, B_o)$. An arbitrary magnetization state trajectory (blue), ending on an example recoil line (black) with associated remanence $B'_{r,i}$ and corner-point $B_{c,i}$ is also drawn. The dashed lines are other possible recoil lines.

where $B_m$ and $B_g$ are the magnetic flux densities in the magnet and the air-gap respectively. The magnetic field intensities are denoted by $H_g$ and $H_m$. In the air-gap, $H_g$ and $B_g$ are related through the permeability of free space $\mu_0$. Parameter $k_1$ is known as the flux leakage coefficient which describes how much of the flux going through the magnet actually arrives in the air-gap. It compensates for the effects of flux leakage and fringing flux. Parameter $k_2$ is the loss factor, which accounts for the potential magnetomotive force loss due to unwanted magnetization of the steel components.

The behavior of PMs in a magnetic circuit is usually described by a load-line [11]. For the magnetic circuit of Fig. 1, the load-line equation can be written by combining (1), (2) and (3) as:

$$B_m = -\mu_0 \left(\frac{k_1}{k_2}\right) \frac{A_g L_m}{2 A_m l_g} \left(H_m - \frac{N I_c}{L_m}\right) \quad (4)$$

It relates the field intensity $H_m$ in the magnet to the resulting magnetic flux density $B_m$. In order to determine the magnet's operating point, another such relation is required. This is provided by the magnet's hysteresis characteristic, or *BH* curve, shown in Fig. 2. The magnet operates at a point where the load-line intersects this curve, denoted by $(H_o, B_o)$.

Demagnetization of the magnet can be caused by a changing circuit reluctance (slope of the load-line) or an applied magnetizing current (position of the load-line), both of which change the magnet's operating point. An important feature of the *BH* curve is the knee-point, as denoted in Fig. 2. Shifting the operating point past this point will yield a permanent demagnetization and subsequent operation over a recoil line. There exist a continuum of these recoil lines within the major *BH* curve, each of which constitutes a possible level of magnetization with its own associated remanence $B'_{r,i}$. The point where the recoil line intersects the major *BH* curve is denoted by corner-point $B_{c,i}$. The recoil lines are linear with a slope equal to the recoil permeability, noted by relative permeability value $\mu_{rec}$ [11].

## 3. Magnetization State Tuning

The previous section elaborated on modeling the *Tunable Magnet*. This section introduces a magnetization state tuning method based on this model. We assume that the air-gap $l_g$ of the TM actuator from Fig. 1 is variable but accurately known. This is often the case for an actuator, where the position of the mover is measured to be used in motion feedback control. The proposed tuning process is visualized in Fig. 3. It is summarized as follows: if a new level of air-gap flux density $B_g^{set}$ is required, the system predicts at what recoil-line the magnet needs to operate to achieve this for a given $l_g$. If this recoil-line is higher than the present one, the magnet is saturated by a saturating voltage pulse on the coil. If not, the magnet is immediately tuned to the correct recoil line, where the demagnetizing voltage pulse is generated by an air-gap flux feedback controller. The point $(H_o, B_o)$ where the magnet has to operate to yield a certain required air-gap flux density $B_g^{set}$ is calculated using (1), (2) and (3):

$$B_o = k_1 \frac{A_g}{A_m} B_g^{set} \quad (5)$$

$$H_o = -2\, k_2 \frac{l_g}{L_m \mu_0} B_g^{set} \quad (6)$$

The recoil-line that corresponds to this operating point, identified by its remanent magnetization is:

$$B'_{r,i} = B_o - \mu_{rec}\, \mu_0\, H_o \quad (7)$$

During demagnetization, the operating point is shifted over the major *BH* curve in the 2nd quadrant. Therefore, the demagnetization controller reference is equal to corner-point $B_{c,i}$ i.e., the point where the desired recoil line intersects the major hysteresis loop. Demagnetizing to this point guarantees subsequent operation over the correct recoil line. An example magnetization state trajectory, corresponding to an arbitrary tuning cycle starting from $B_m = 0$ is depicted in Fig. 2 in blue.

## 4. Controller Design

To understand the dynamics of the system that the controller is working on, the effect of a coil voltage $U_c$ on the air-gap flux density $B_g$ in the TM actuator is derived. This relation can be described by the following transfer function:

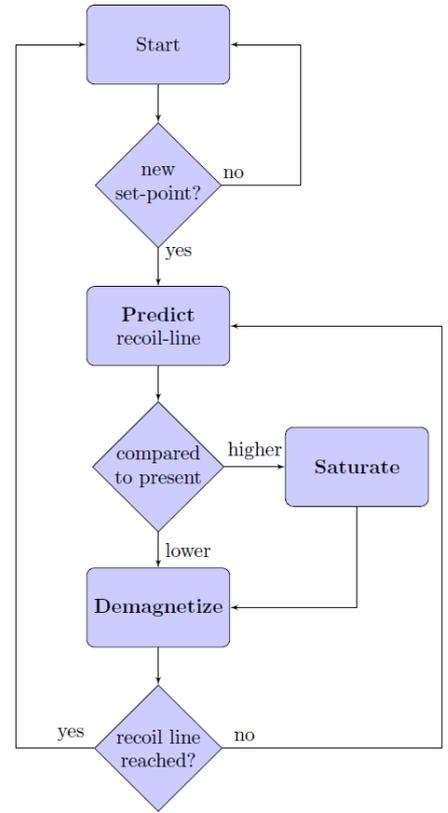

**Figure 3:** Magnetization State tuning method consisting of 3 steps: **Predict, Saturate and Demagnetize**.

$$G(s) = \frac{B_g(s)}{U_c(s)} = \frac{G_0(\mu, l_g)}{\frac{L(\mu, l_g)}{R} s + 1} \quad (8)$$

where $L(\mu, l_g)$ and $G_0(\mu, l_g)$ are the non-linear inductance and the DC gain of the system. Both are dependent on the magnet permeability $\mu = \frac{dB_m}{dH_m}$, which in term depends on the magnetization state. The plant to be controlled is thus non-linear with a gain that depends on the AlNiCo operating point.

The proposed demagnetization controller has a structure as shown in Fig. 4. Since it uses air-gap flux feedback, the reference input $B_{c,i}$ is first translated to the associated air-gap value using (1). Since demagnetization along the major *BH* curve in the second quadrant is irreversible, it is imperative that the controlled system has zero overshoot. Otherwise, the correct corner-point will be passed, and the magnet will end up operating on an incorrect recoil-line. This also requires the step response to have zero steady state error. A simple PI controller will add 90° of phase, ensuring a damped step response without overshoot. The integrator gives high gain at low frequencies, therefore minimizing steady-state error. The tuned PI controller has the following form:

$$C(s) = k_p + \frac{k_i}{s} \quad (9)$$

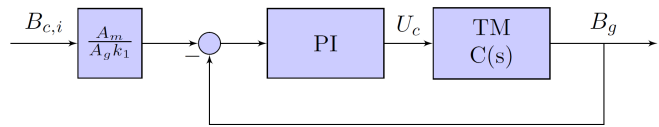

where $k_p = 2.07\text{ V T}^{-1}$ and $k_i = 150\text{ V T}^{-1}\text{s}^{-1}$. These controller gains give good performance even with the non-linear plant $G(s)$. A typical step-response of the system is shown in Fig. 7 with the reference input $B_{c,i}$ denoted by the dotted red line.

**Figure 4:** PI demagnetization controller. Reference $B_{c,i}$ is first transformed to the associated air-gap flux density.

## 5. Experimental results

In this section, we present the hardware implementation of the *Tunable Magnet* and the experimental verification of the proposed magnetization state tuning method. Figure 5 shows the experimental setup with the TM implemented in a reluctance actuator topology, similar to Fig. 1. As PM material we use AlNiCo 5, which is the most common variation with the highest remanent magnetization. The pole pieces and keeper bar are made of soft steel (st. 37). We measure air-gap flux density by combining a hall sensor and a sense coil, to accurately capture both DC and AC values. Both sensors are placed in a 3D printed fixture of 1.00 mm thickness to accurately fix the nominal air-gap width (Fig. 5). The air-gap can then be varied by moving the keeper bar on a manual linear stage (resolution 10 μm). Magnetizing current is provided by a linear power amplifier and measured by the voltage drop over a sense resistor.

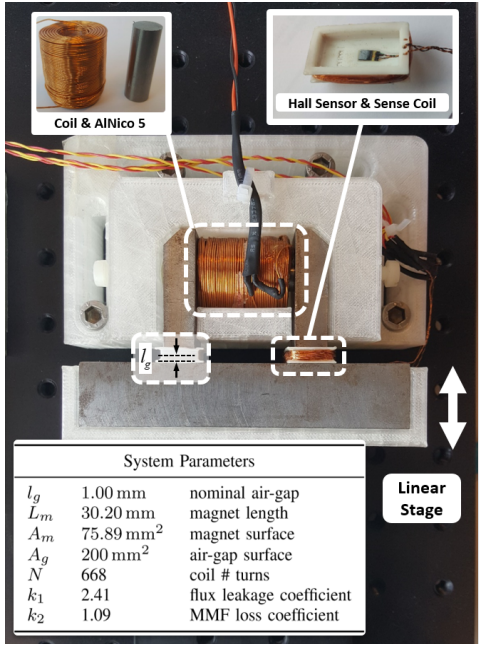

**Figure 5:** TM measurement setup with parameter values. The keeper bar is mounted on a manual linear precision stage to vary the air-gap.

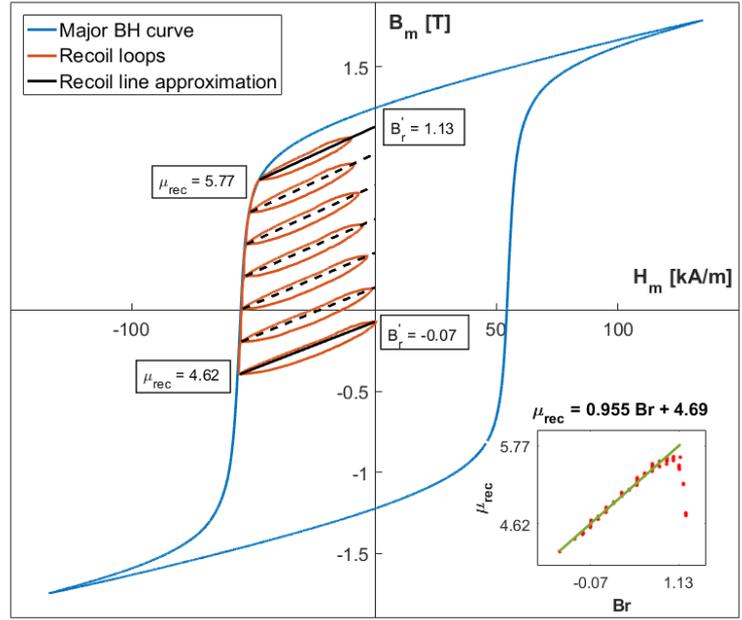

**Figure 6:** Estimate of the major *BH* curve with recoil-loops using measurements of $B_g$ and $I_c$ together with (1) and (2). The inset shows the distribution of permeabilities $\mu_{rec}$, for recoil-loops at different remanent magnetizations $B'_r$ in the 2nd quadrant.

### 5.1. BH curve *Characterization:*

To use the earlier derived model, we first need to identify coefficients $k_1$ and $k_2$. This is done using magnetic FEA in COMSOL, with a detailed 3D model of the TM reluctance actuator. Next, the AlNiCo 5 is driven through its major *BH* curve at low frequency, as well as several recoil-loops. Current and air-gap measurements from this experiment are then used to estimate $H_m$ and $B_m$ inside the magnet using (1),(2) and (3). This results in the *BH* characteristic shown in Fig. 6. Note that the earlier described recoil-lines are actually small loops, of which the recoil-permeability $\mu_{rec}$ and associated remanence $B'_{r,i}$ can be approximated by connecting the corners with a straight line. The inset of Fig. 6 shows the relation between recoil-line position, denoted by $B'_r$ and recoil-line permeability $\mu_{rec}$. Over de second quadrant of the *BH* curve this relation can be approximated as linear:

$$\mu_{rec} = 0.955\, B'_r + 4.69 \tag{10}$$

### 5.2 Magnetization State Tuning Measurements:

Combining (7) and (10) and solving numerically for $B'_r$ gives the required remanent magnetization for a desired $B_g^{set}$. Intersecting the associated recoil-line (7) with the measured major *BH* curve gives a value for corner-point and demagnetization controller reference input $B_c$. With this information the magnetization state tuning method of Fig. 3 can now be implemented.

Figure. 7 shows the air-gap flux density $B_g$ over time for a typical tuning cycle, and Fig. 8 the corresponding magnetization state trajectory in terms of $H_m$ and $B_m$. Starting from $B \approx 0$ T, the magnet is driven to saturation $B_{sat}$ and demagnetized to the correct corner-point $B_c$. When the coil voltage is removed the current decays, and the magnetization approaches the desired operating point $B^{set}$. During the saturation pulse, the BH trajectory does not exactly follow the major *BH* curve. This is due to a dynamic effect called loop-widening [12], caused by induced eddy currents. It does not, however, influence the operating point reached.

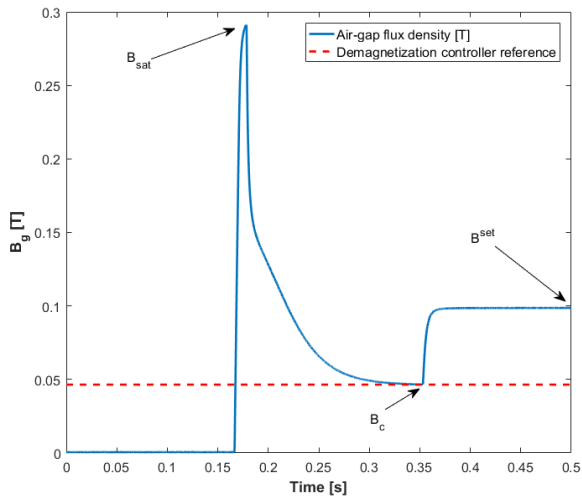

**Figure 7:** Measured $B_g$ response during a typical tuning cycle, with annotated saturation point, corner point, and flux density set-point.

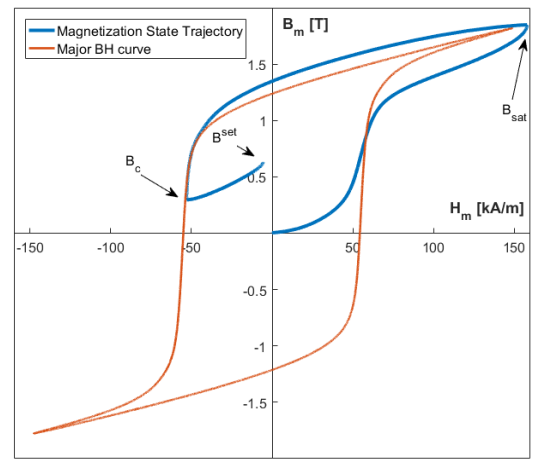

**Figure 8:** Measured magnetization state trajectory for a typical tuning cycle, with annotated saturation point, corner point and flux density set-point.

With the given demagnetization controller, the time it takes to do a complete tuning cycle is in the order of 500 ms. The exact value depends on $B^{set}$, $l_g$ and system geometry. The tuning speed can be increased by modifying the controller (9) if enough power supply voltage is available. It is however ultimately limited by dynamic effects such as loop widening.

To evaluate the robustness of the TM implementation, we recorded 44 tuning cycles for different air-gap flux density set-points $B_g^{set}$ at different air-gaps $l_g$. Tables 1 and 2 show the achieved accuracy, in terms of the MEA (mean absolute error), and the precision ($3\sigma$) for $l_g = 1.00$ mm and $l_g = 1.20$ mm. Note that the air-gap is constant during tuning. In general, it can be concluded that the tuning error increases with increasing values of $B_g^{set}$. This is caused by the fact that the absolute value of the error in predicting the operating point is proportional to $B_g^{set}$ according to (5) and (6). The comparatively large error at $B_g^{set} = 0.175$ T can be explained by the added inaccuracy in approximating the distribution of $\mu_{rec}$ as linear. For recoil lines approaching the major *BH* curve, this approximation becomes invalid, as shown in Fig. 6. Comparing the results for both air-gaps, we see that the error increases with increasing $l_g$. This mainly due to the limited accuracy of the used magnetic circuit model. For example, the flux leakage and fringing effects will increase with increasing air-gap i.e., $k_1$ is air-gap dependent. The more the air-gap deviates from the nominal value at which the model was identified, the larger the operating point prediction error and consequently the tuning error.

**Table 1,2:** Tuning performance results over n = 44 tuning cycles for each $B_g^{set}$ at different $l_g$.

| $B_g^{set}$ | Realized values | | $l_g = 1.00$ mm | $B_g^{set}$ | Realized values | | $l_g = 1.20$ mm |
|---|---|---|---|---|---|---|---|
| | Mean | MAE* | Precision ($3\sigma$) | | Mean | MAE* | Precision ($3\sigma$) |
| 0.175 T | 0.167 T | 7.61 mT | 0.67 mT | 0.175 T | 0.159 T | 15.86 mT | 0.11 mT |
| 0.150 T | 0.146 T | 4.30 mT | 0.10 mT | 0.150 T | 0.140 T | 9.88 mT | 0.60 mT |
| 0.125 T | 0.122 T | 2.62 mT | 0.11 mT | 0.125 T | 0.118 T | 7.48 mT | 0.19 mT |
| 0.100 T | 0.099 T | 1.37 mT | 0.13 mT | 0.100 T | 0.094 T | 6.04 mT | 0.40 mT |
| 0.075 T | 0.074 T | 0.50 mT | 0.11 mT | 0.075 T | 0.070 T | 5.14 mT | 0.53 mT |
| 0.050 T | 0.050 T | 0.16 mT | 0.10 mT | 0.050 T | 0.045 T | 4.54 mT | 0.54 mT |
| 0.025 T | 0.026 T | 0.72 mT | 0.14 mT | 0.025 T | 0.021 T | 4.10 mT | 0.56 mT |
| 0.000 T | 0.001 T | 1.07 mT | 0.11 mT | 0.000 T | −0.004 T | 3.81 mT | 0.15 mT |

\* = Mean Absolute Error

## 6. Conclusion

In this paper, we have presented a way of implementing a *Tunable Magnet*, which enables it to be used in actuation applications. It was shown that robust magnetization state tuning for a variable but known air-gap could be achieved. This is accomplished by predicting the right operating point for the AlNiCo 5 PM, based on measured *BH* data and a magnetic circuit model. An air-gap flux feedback controller is then used to generate the demagnetizing current such that the magnet approaches this operating point. If necessary, the magnet is saturated first. Using this method we have achieved magnetization state tuning with a maximum error of 15.86 mT and a minimum precision of 0.67 mT, for air-gap flux density set-points in the range of $0 \leq B_g^{set} \leq 0.175$ T. This was done for air-gaps in the range of 1.00 mm $\leq l_g \leq 1.20$ mm. With the obtained tuning accuracy, actuator heating during static periods is almost eliminated. Only a small bias current is needed to compensate for the tuning error. Tuning times are in the order of 500 ms per cycle, but can be increased by modifying the demagnetization controller. Ultimately, the tuning speed is limited by the magnetization dynamics of the AlNiCo PM used.

During the tuning cycle, the air-gap was fixed. The next step toward a functional TM actuator is to evaluate the performance of the proposed tuning method in a situation where the air-gap is actually varying during the tuning cycle. Also, the accuracy of the operating point prediction step can be increased by using a more accurate magnetic circuit model, and a better fit for the recoil permeability distribution. The applicability of the TM can be improved by changing the demagnetization controller to increase the tuning speed. Exact characteristics of the quasi-static applications in which the TM increases energy efficiency also need to be investigated and precisely defined.